\documentclass[reprint, amsmath, amssymb, aps,pra]{revtex4-2}
\usepackage{xcolor}
\numberwithin{equation}{section}

\usepackage{xcolor}
\usepackage{xfrac}
\usepackage[
  colorlinks=true,
  linkcolor=blue,
  citecolor=blue,
  urlcolor=blue
]{hyperref}
\usepackage{mathrsfs}
\usepackage{graphicx}
\usepackage{bm}
\usepackage{orcidlink}
\usepackage{braket}
\usepackage[normalem]{ulem}

\usepackage{academicons}
\newcommand{\N}{\text{N}}

\newcommand{\cnn}{Universit\'e Paris-Saclay, Centre de Nanosciences et de Nanotechnologies, CNRS, 10 Boulevard Thomas Gobert, 91120, Palaiseau, France}

\begin{document}

\title{Nonclassicality of multi-photon-added cat states}

\author{Jhordan Santiago\,\orcidlink{0009-0001-9265-9827}}
 \email{jhordansantiago03@gmail.com}
\affiliation{Unidade Acadêmica de Física, Universidade Federal de Campina Grande, 10071, 58429-900, Campina Grande, Paraíba, Brazil}

\author{Petr Steindl\,\orcidlink{0000-0001-9059-9202}}
\email{petr.steindl@cnrs.fr}
\affiliation{\cnn}

\date{\today}
\begin{abstract}
Multi-photon-added cat states are constructed by repeatedly applying the creation operator to a cat state. We study in detail their photon-number distribution, $Q$ parameter, squeezing properties, and Wigner function. We show that photon addition induces a $\pi$ phase shift in the original parity configuration whenever an odd number of photons is added, reflected as swapped vanishing probabilities and phase space displacements at the origin. Remarkably, the same process drives these states into a sub-Poissonian regime regardless of the relative phase between their coherent state components, making them valuable resources for quantum imaging, at the cost of losing quadrature squeezing, but gaining amplitude-squared one. We also discuss how these states can be generated using existing hardware.
\end{abstract}

\maketitle
\section{Introduction}  
Nonclassical states of light have been extensively studied for their remarkable properties, which make them valuable resources for quantum computing~\cite{Knill2001,Raussendorf2001,AghaeeRad2025,Maring2024, Lloyd1999} and high-precision metrology \cite{Zurek2001}. Among these states, cat states $|\alpha\rangle + |-\alpha\rangle$ (unnormalized), defined as coherent superpositions of out-of-phase coherent states $|\alpha\rangle$~\cite{Dodonov1974,Yurke1986}, have attracted attention as a resource for breeding Gottesman–Kitaev–Preskill states \cite{Winnel2024,Larsen2025} and for implementing error correction codes \cite{Minganti2016,Bergmann2016,Ofek2016,Hastrup2022}. These states, even under strong mixing, retain coherence, which is important for fault-tolerant quantum information processing~\cite{Yang2025}.

Non-Gaussian gates, such as photon addition~\cite{Kim2008,Marco2010} and subtraction~\cite{Wenger2004} from a photonic field, enable
the enhancement of thecement of the nonclassicality of the photonic state. Combining these operations sequentially, in principle, allows for the full engineering of the photon statistics of a photonic state~\cite{Biagi2022}. This is based on the commutation relation $[a,a^\dagger]=\mathbf{1}$, where photon-addition (subtraction) implements the annihilation (creation) operator $a$ ($a^\dagger$)~\cite{Parigi2007,Zavatta2008,Parigi2009}.

Applied to cat states, it is well-established that an odd number of photon subtractions converts an even cat state into an odd one \cite{Barnett2018, Puri2017}. Since photon addition transforms any Gaussian state into a non-Gaussian one \cite{Hertz2023}, a natural extension is to investigate the configuration in which the field is prepared in an excited cat state. Photon-added even and odd cat states were introduced early on in \cite{Dodonov1996,Xin1996}, where their sub-Poissonian and (first-order) non-squeezed character was already pointed out in substantial detail, with the single-photon case later rediscovered in \cite{Arman2021}. However, these studies restrict themselves to a single-excitation or to specific relative phases $\phi$ of the cat state. To the best of our knowledge, a general analysis of multi-photon-added cat states, with arbitrary addition of $m$-photons and the relative phase $\phi$, has not been reported. It is therefore compelling to explore this regime: although even and odd cats are among the most mathematically and physically interesting cases - being eigenstates of the parity operator \cite{Dodonov1974} and representations of the crystallographic $C_2$ point group \cite{Castanos1995} - the Yurke-Stoler state \cite{Yurke1986}, Poissonian and squeezed, is not covered by such results. 

Motivated by this gap, in this Letter, we describe multi-photon-added cat states with arbitrary relative phase $\phi$ using a single formalism, investigate their nonclassical properties, and discuss how they can be produced using current hardware. We show that photon
addition induces a $\pi$ phase shift in the parity configuration
of the original superposition, a feature clearly visible in
both the photon-number distribution and in the phase
space. Although quadrature squeezing and sub-Poissonian behavior
in cat states depend on the relative phase $\phi$ between their
coherent components \cite{Gerry1997}, we find that multi-photon-added cat states are always sub-Poissonian but never
quadrature-squeezed, independently of the phase. Interestingly, the same process that suppresses first-order squeezing simultaneously gives rise to amplitude-squared (Hillery-type) squeezing \cite{Hillery1987a}, revealing a more subtle redistribution of quantum fluctuations beyond the standard quadrature picture.
\section{Photon-added cat state}\label{2.2}

We define (in atomic units $\hbar=\omega=1$) the $m$ photon-added cat state as
\begin{align}
|\alpha, -\alpha, m\rangle = \text{N}_m^{-1/2}\big(a^{\dagger m}|\alpha\rangle + e^{i\phi}a^{\dagger m}|-\alpha\rangle\big),
\label{eq:cat_state}
\end{align}
with normalization
\begin{align}
\text{N}_m = 2m!\Big[ \mathscr{L}_m(-|\alpha|^2) + \mathscr{L}_m(|\alpha|^2)e^{-2|\alpha|^2}\cos\phi \Big],
\label{norm}
\end{align}
where $\mathscr{L}_m(x)$ is the Laguerre polynomial of order $m$ \cite{Gradshteyn2014}, and $\phi$ is the relative phase between the coherent states.

For $\phi=0,\ \pi/2$ and $\pi$, Eq. \eqref{eq:cat_state} yields the photon-added even, Yurke-Stoler and odd cat states, respectively. Note that these states differ from a superposition of displaced number states \cite{Laiho2012,Lam2025}, since the displacement and creation operators do not commute. In the limit $\alpha\rightarrow 0$ ($m\rightarrow0$), the state approaches the number (cat) state.

Photon-added cat states display rich nonclassical behavior. For odd $m$, the superposition gains a $\pi$ phase translation, reflecting a parity change of its number components. This follows from the fact that the cat state is an eigenstate of the squared annihilation operator $a^{2}$~\cite{Gerry1997}, hence, after an even number of photon subtractions, the initial cat state is exactly recovered; this property has proven advantageous in quantum error correction~\cite{Mirrahimi2014,Guillaud2023}. In contrast, an even number of photon additions restores only the parity structure, not the full state. The reason is that the creation operator acting on a coherent state generates two contributions: one proportional to $|\alpha\rangle$ and another corresponding to a displaced $m$-excitation. In particular,
\begin{align} |\alpha,-\alpha,m\rangle &= \text{N}_m^{-1/2} \sum_{p=0}^{m} \binom{m}{p} \sqrt{p!} \Big[ (\alpha^*)^{m-p} D(\alpha)\nonumber\\&\quad+ e^{i\phi}(-\alpha^*)^{m-p} D(-\alpha) \Big]|p\rangle,\label{eq:photon-added-dis} \end{align}
where $D(\alpha)$ is the displacement operator. Hence, after two photon additions, the state does not coincide with the original cat, even though its parity is recovered.

\section{Nonclassicality}\label{3.3}
\subsection{Oscillatory photon-number distribution}
In a radiation field prepared in a photon-added cat state, the probability of detecting $n$ photons is given by
\begin{align}
P(n) &=|\langle n|\alpha,-\alpha,m\rangle|^2\nonumber\\
&=\dfrac{e^{-|\alpha|^2}n!}{\text{N}_m\big[(n-m)!\big]^2}|\alpha|^{2(n-m)}
\left[2+2(-1)^{n-m}\cos\phi\right],\label{prob}
\end{align}
    for $m\leq n$, while $P(n<
    m)=0$. 
    
\begin{figure}[htbp]
\includegraphics[width=\linewidth]{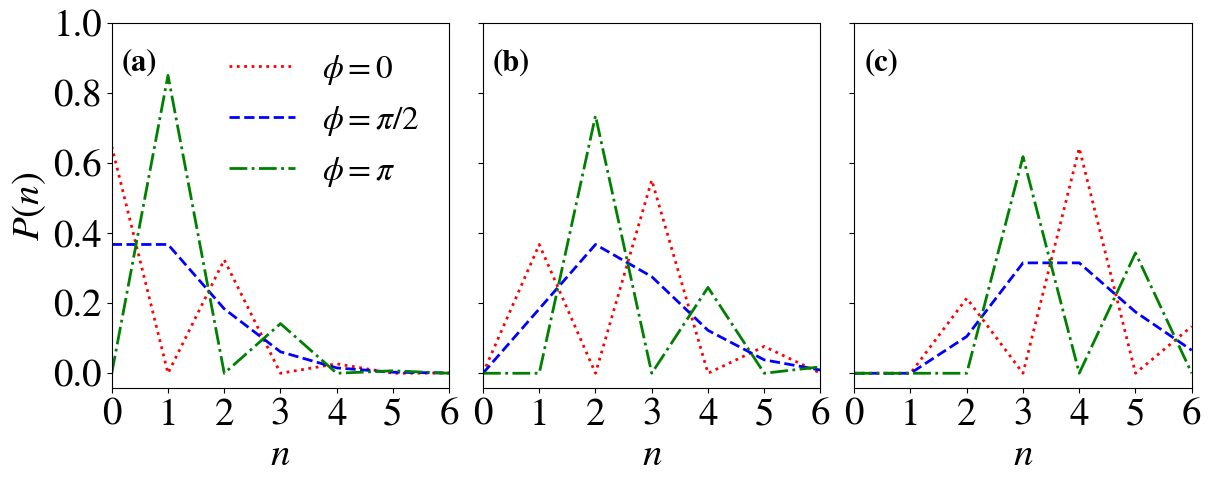}
\caption{Photon-number distributions of the (a) cat state with $\alpha=1$ and relative phase $\phi$ (different color), (b) after its single-photon addition, and (c) after two-photon addition, restoring the original parity.}
\label{fig1}
\end{figure}

The oscillatory behavior of $P(n)$ in Fig.~\ref{fig1} arises from the quantum interference between its number components, which periodically enhances or suppresses the probability amplitude \cite{Schleich2001}.  
For $\phi=0$ (even cat), constructive interference occurs whenever $n-m$ is even, while destructive interference leads to $P(n)=0$ for odd $n-m$.  
For $\phi=\pi$ (odd cat), the situation is reversed.  
The Yurke-Stoler case ($\phi=\pi/2$) satisfies $\cos\phi=0$, eliminating the interference term and yielding a Poisson-like, non-oscillatory distribution.

\subsection{Sub-Poissonian character of the field}

The statistical nature of the field can be quantified using the $Q$ parameter \cite{Fano1947,Mandel1979}, defined as
\begin{align}
Q=\frac{\langle a^{\dagger 2}a^2\rangle}{\langle a^\dagger a\rangle^2},
\label{q}
\end{align}
where Poissonian statistics correspond to $Q=1$, sub-Poissonian to $Q<1$, and super-Poissonian to $Q>1$.

Using the antinormal-ordered identity $\langle a^x a^{\dagger x}\rangle = \text{N}_{m+x}/\text{N}_m$ \cite{Ren2014} and the normalization factor in Eq.~\eqref{norm}, Eq.~\eqref{q} can be written as
\begin{align} Q = \frac{\text{N}_{m+2}-4\text{N}_{m+1}+2\text{N}_m}{\text{N}_{m+1}-\text{N}_m} - \frac{\text{N}_{m+1}}{\text{N}_m}.\label{mandel} \end{align}

\begin{figure}[htbp]
  \centering
\includegraphics[width=\linewidth]{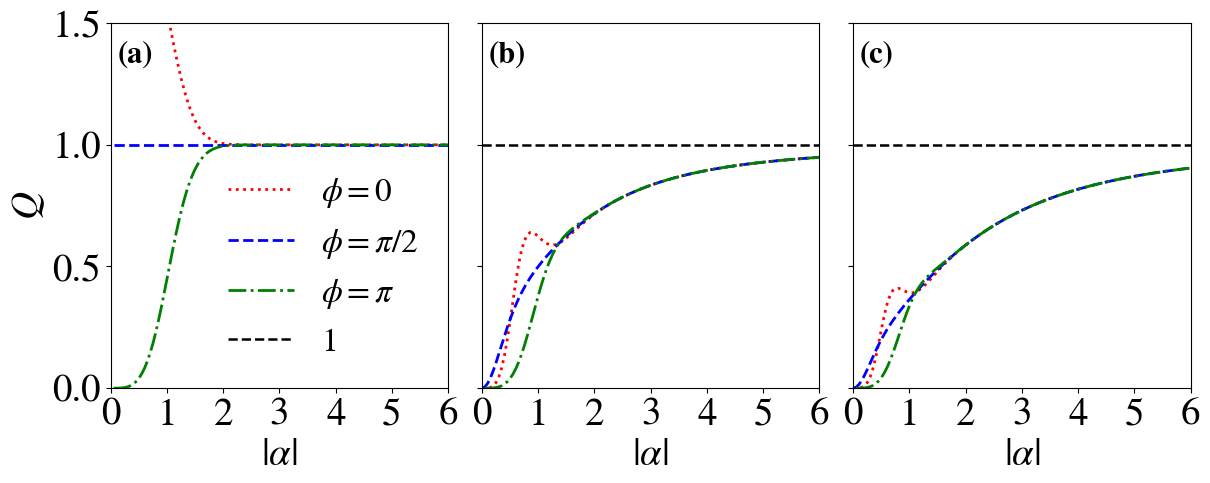}
\caption{The $Q$ parameter as a function of cat state strength $|\alpha|$  and different $\phi$ (color encoded) evaluated for (a) the cat state, (b) the single-photon-added cat state, and (c) the two-photon-added cat state.}
\label{fig2}
\end{figure}

Fig. \ref{fig2} shows that photon addition converts both the Poissonian Yurke-Stoler state and the super-Poissonian even cat state into sub-Poissonian ones ($Q<1$). This effect becomes stronger with increasing $m$, and the distribution approaches the limit $Q\to 0$ when $m\gg|\alpha|$, where the added-photon component dominates and the photon-number fluctuations are strongly suppressed. In the opposite limit, when $|\alpha|\gg m$, the state approaches the Poisson distribution.

The odd cat state is already sub-Poissonian and remains so under photon addition. Notably, $Q<1$ holds for all $\phi\in[0,\pi]$, meaning that the sub-Poissonian character of photon-added cat states is relative phase-independent. This robustness makes them particularly attractive for quantum-enhanced imaging, where reduced photon-number noise is advantageous \cite{Berchera2019}. Finally, we emphasize that $Q>1$ does not imply classicality, and that (sub/super)-Poissonian behavior should not be confused with (anti-)bunching, since a quantum field may exhibit any combination of these properties \cite{Singh1983,Zou1990,Emary2012}.

\subsection{Squeezing properties}

In this section, we examine lower and higher-order squeezing of the photon-added cat states. Squeezing is a purely quantum effect in which the fluctuations of a field quadrature are reduced below the vacuum level, at the expense of increased fluctuations in the conjugate quadrature \cite{Fox2006,WallsMilburn2025}. This phenomenon is necessary to high-precision metrology such as gravitational wave detectors \cite{Chua2014,Heinze2022,Nishino2024,Jia2024} or quantum imagining \cite{Lawrie2019}. Therefore, if photon-added cat states exhibit any form of squeezing, they could serve as valuable resources in such platforms.

The first-order field quadrature, where $\theta$ is a global phase rotation in phase space, is defined as\begin{equation}
    x_\theta = \frac{a e^{i\theta} + a^\dagger e^{-i\theta}}{2},
\end{equation}
and quadrature squeezing occurs when $(\Delta x_\theta)^2 < 1/4$, with
$(\Delta x_\theta)^2 = \langle x_\theta^2 \rangle - \langle x_\theta \rangle^2$. The first moment evaluates the $m$-photon-added cat states to
\begin{align}
    \langle x_\theta \rangle =
    -\frac{|\alpha|e^{-2|\alpha|^2}\mathscr{L}_m^{(1)}(|\alpha|^2)
    \sin\phi\sin(\theta-\varphi)}
    {\mathscr{L}_m(-|\alpha|^2)+\mathscr{L}_m(|\alpha|^2)e^{-2|\alpha|^2}\cos\phi},\label{var}
\end{align}
where $\mathscr{L}_m^{(y)}(x)$ is the associated Laguerre polynomial \cite{Gradshteyn2014} and $\alpha=|\alpha|e^{i\varphi}$. Since Eq. \eqref{var} is proportional to $\sin\phi$, it vanishes for all $\phi \in [0,\pi]$ except for $\phi = \pi/2$. The second moment is
\begin{align} \langle x_\theta^2 \rangle &= \frac{m!|\alpha|^2\cos[2(\theta-\varphi)]}{\text{N}_m} \big[ \mathscr{L}_m^{(2)}(-|\alpha|^2) \nonumber\\&\quad+e^{-2|\alpha|^2}\mathscr{L}_m^{(2)}(|\alpha|^2)\cos\phi \big] \nonumber\\ &\quad + \frac{(m+1)!}{\text{N}_m} \big[ \mathscr{L}_{m+1}(-|\alpha|^2) \nonumber\\&\quad +e^{-2|\alpha|^2}\cos\phi\mathscr{L}_{m+1}(|\alpha|^2) \big] -\frac{1}{4}. \end{align}

To assess whether the photon-added cat state exhibits quadrature squeezing, an asymptotic analysis is required (where $O(x)$ denotes the expansion order).  In the small-amplitude limit $|\alpha|\ll1$ and for any fixed $m>0$, expanding the Laguerre polynomials at the origin gives $\mathscr{L}_m(\pm|\alpha|^2)\simeq1\mp m|\alpha|^2$ and $\mathscr{L}_m^{(2)}(0)=(m+2)(m+1)/2$. To leading order, one finds $(\Delta x_\theta)^2\simeq m+3/4$, which is well above the vacuum value $1/4$, signaling no squeezing. In the opposite limit $|\alpha|\gg1$, the interference terms proportional to $e^{-2|\alpha|^2}$ are exponentially small, and the dominant contribution scales as $\langle x_\theta^2\rangle\simeq2|\alpha|^2\cos^2(\theta-\varphi)$. For a generic angle the variance grows as $O(|\alpha|^2)$, again exceeding the vacuum noise. Only at the special quadrature $\cos(\theta-\varphi)=0$ does the $O(|\alpha|^2)$ term cancel; however, the remaining corrections are $O(1)$ or exponentially small in $|\alpha|^2$, so any residual squeezing becomes negligible for large amplitudes. Therefore, photon-added cat states do not exhibit quadrature squeezing.  
This is noteworthy because both the even and the Yurke-Stoler cat can show squeezing before photon addition \cite{Gerry1997}, yet their photon-added counterparts lose this property.  
Together with the results of Ren et al. \cite{Ren2014} for the photon-added compass state, these findings suggest a strong conjecture: non-squeezing appears to be a generic feature of photon-added cyclic coherent states \cite{Castanos1995}.

\begin{figure}[htbp]    \centering\includegraphics[width=\columnwidth]{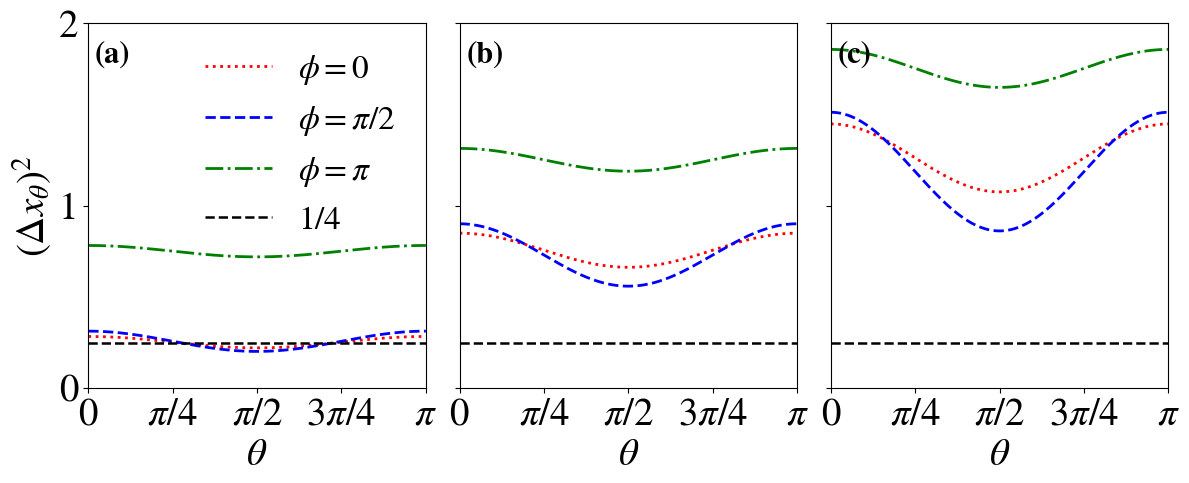} \caption{Variance $(\Delta x_\theta)^2$ as a function of $\theta$, for weak cat state with $\alpha=0.25$ and relative phase $\phi$ (color encoded), for the (a) non-excited, (b) single-excited and (c) double-excited cat state. }\label{fig4}\end{figure}

To demonstrate that the photon-added cat states are not quadrature squeezed, we
plot in Fig.~\ref{fig4} $(\Delta x_\theta)^2$ as a function of $\theta$ for a fixed weak amplitude $\alpha = 0.25$. In this regime, both the even and the Yurke-Stoler cat states exhibit squeezing, so one can naively expect some degree of squeezing after the photon addition. However, no squeezing is observed in Fig.~\ref{fig4}b and c after adding one or two photons, respectively - in agreement with our previous asymptotic analysis.

\begin{figure}[htbp]
    \centering
\includegraphics[width=\columnwidth]{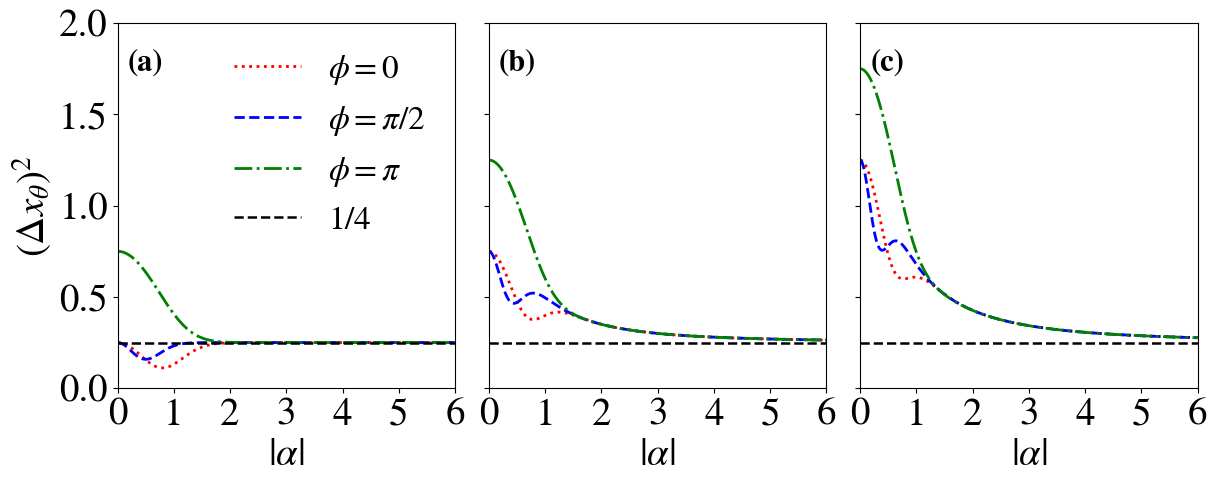}
    \caption{Variance $(\Delta x_\theta)^2$ as a function of $|\alpha|$, with fixed phase $\theta+\varphi=\pi/2$, for (a) the cat state, (b) a single-photon-added cat, and (c) a two-photon-added cat.}
    \label{fig3}
\end{figure}

In Fig.~\ref{fig3}, we present a numerical evaluation of the quadrature variance for the photon-added cat state, setting the quadrature with minimal variance, satisfying $\theta+\varphi=\pi/2$ (see Fig.~\ref{fig4}). The plot shows that, as the number of added photons increases, the state becomes progressively less squeezed. In the limit $|\alpha|\rightarrow\infty$ the variance approaches the vacuum level, corresponding to the asymptotic approach to a coherent state.

Now, we turn our attention to higher-order squeezing, in particular the Hillery-type amplitude-squared squeezing \cite{Hillery1987a,Hillery1987b}. The squared quadrature is defined as
\begin{align}
y_\theta = \frac{a^2 e^{i\theta} + a^{\dagger 2} e^{-i\theta}}{2},\label{xs}
\end{align}
where second-order squeezing occurs when $Y(\theta)\equiv(\Delta y_\theta)^2-|\langle a^\dagger a+1/2\rangle|<0$, with
$(\Delta y_\theta)^2 = \langle y_\theta^2 \rangle - \langle y_\theta \rangle^2$. Omitting the algebraic manipulation, the first and second moments of Eq. \eqref{xs} are given by
\begin{align}
    \langle y_\theta \rangle
    &= 2\frac{m! |\alpha|^2}{\N_m}
    \cos(2\varphi + \theta)\nonumber\\
    &\quad
   \times \big[
        \mathscr{L}_m^{(2)}(-|\alpha|^2)
        + e^{-2|\alpha|^2} \mathscr{L}_m^{(2)}(|\alpha|^2)\cos\phi
    \big],
\end{align}

\begin{align}
     \langle y_\theta^2\rangle&=\frac{m! |\alpha|^4}{\N_m}
       \cos(4\varphi + 2\theta)
       \big[
           \mathscr{L}_m^{(4)}(-|\alpha|^2)
           \nonumber\\&\quad+ e^{-2|\alpha|^2}\mathscr{L}_m^{(4)}(|\alpha|^2)\cos\phi
       \big] + \frac{\N_{m+2} - 2\N_{m+1} + \N_m}{2\N_m}.
\end{align}

\begin{figure}[ht]\centering
\includegraphics[width=\columnwidth]{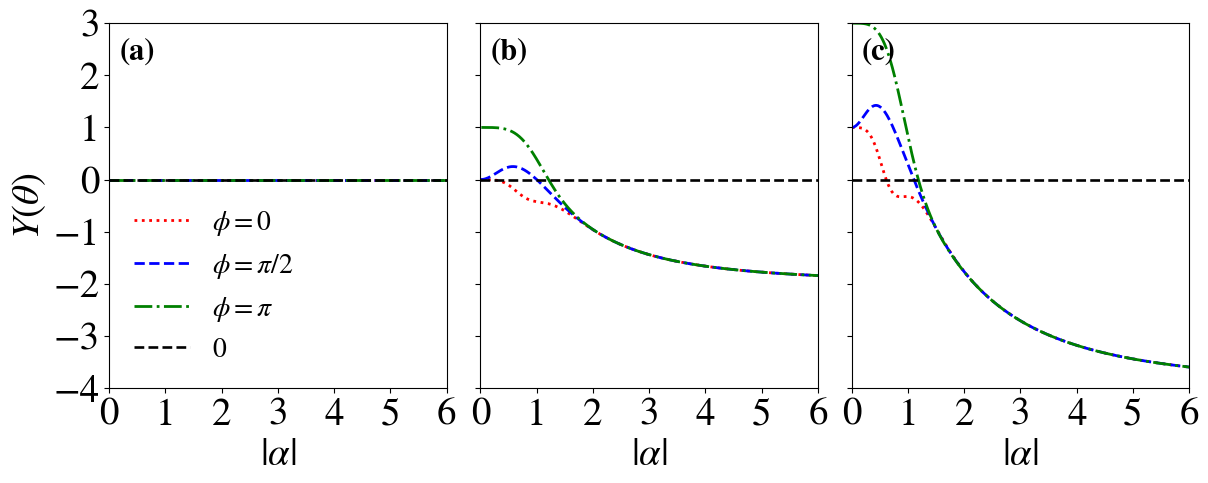}
    \caption{Second-order squeezing $Y(\theta)$ as a function of $|\alpha|$, with fixed $\theta+\varphi=0$, for the cat state (a) before photon-addition, (b) after single-photon addition, and (c) after another photon addition.}
    \label{fig5}
\end{figure}

In Fig.~\ref{fig5}, we present the amplitude-squared squeezing as a function of $|\alpha|$. Since the cat state is an eigenstate of $a^2$, it does not exhibit (as shown in Fig.~\ref{fig5}a), amplitude-squared squeezing \cite{Ahmad2008}; in this sense, it plays a role analogous to that of the coherent state in standard quadrature squeezing \cite{Hillery1987a}. Remarkably, once photons are added (in Fig.~\ref{fig5}b and c), the resulting states not only exhibit second-order squeezing, but the degree of squeezing increases systematically with the number of added photons. This property is particularly noteworthy because, although this type of squeezing has not yet been reported, its detection via homodyne methods has been proposed \cite{Prakash2010,Prakash2012}.

Another interesting property is that the optimal phase space angle becomes $\theta+\varphi=0$, indicating that the axis of the second-order squeezing is rotated by $\pi/2$ with respect to the optimal angle of first-order squeezing. This illustrates that photon addition, and nonlinear operations in general, can dramatically modify the nonclassical properties of cat states. For instance, second-order squeezing has also been reported in cat states after postselected von Neumann measurements \cite{Aishan2022}.
\subsection{Wigner function negativity} 

Negativity regions in the Wigner function are a good indicator of nonclassicality and can be measured by its negative volume \cite{Kenfack2004}. For the photon-added cat state, the Wigner function can be decomposed as
 \begin{align}
W(z) = \text{N}_m^{-1} \Big\{ W_{++} + W_{--} + 2\Re\big[e^{-i\phi} W_{+-}\big] \Big\},
 \label{3.1}
 \end{align}
the first two terms correspond to the unnormalized photon-added coherent states $|\pm\alpha,m\rangle$ \cite{Agarwal1991}, while the third arises from their quantum interference; the latter can be evaluated using the formula \cite{Agarwal1970}\begin{align} W(z)_{+-}=\frac{2}{\pi^{2}} e^{2|z|^{2}} \int d^{2}\beta \langle -\beta |a^{\dagger m}|\alpha\rangle\langle-\alpha|a^m | \beta \rangle e^{2(\beta^{*}z-\beta z^{*})}, \end{align} 
 where $|z\rangle$ is the coherent state, yielding
 \begin{align}
     W_{+-}&=2(-1)^m\pi ^{-1}e^{2|z|^2}e^{-(2z^*+\alpha^*)(2z-\alpha)}\nonumber
     \\&\quad\times\mathscr{L}_m((2z^*+\alpha^*)(2z-\alpha)).\label{3.2}
 \end{align}

\begin{figure*}[htbp]
    \centering    \includegraphics[width=0.8\linewidth]{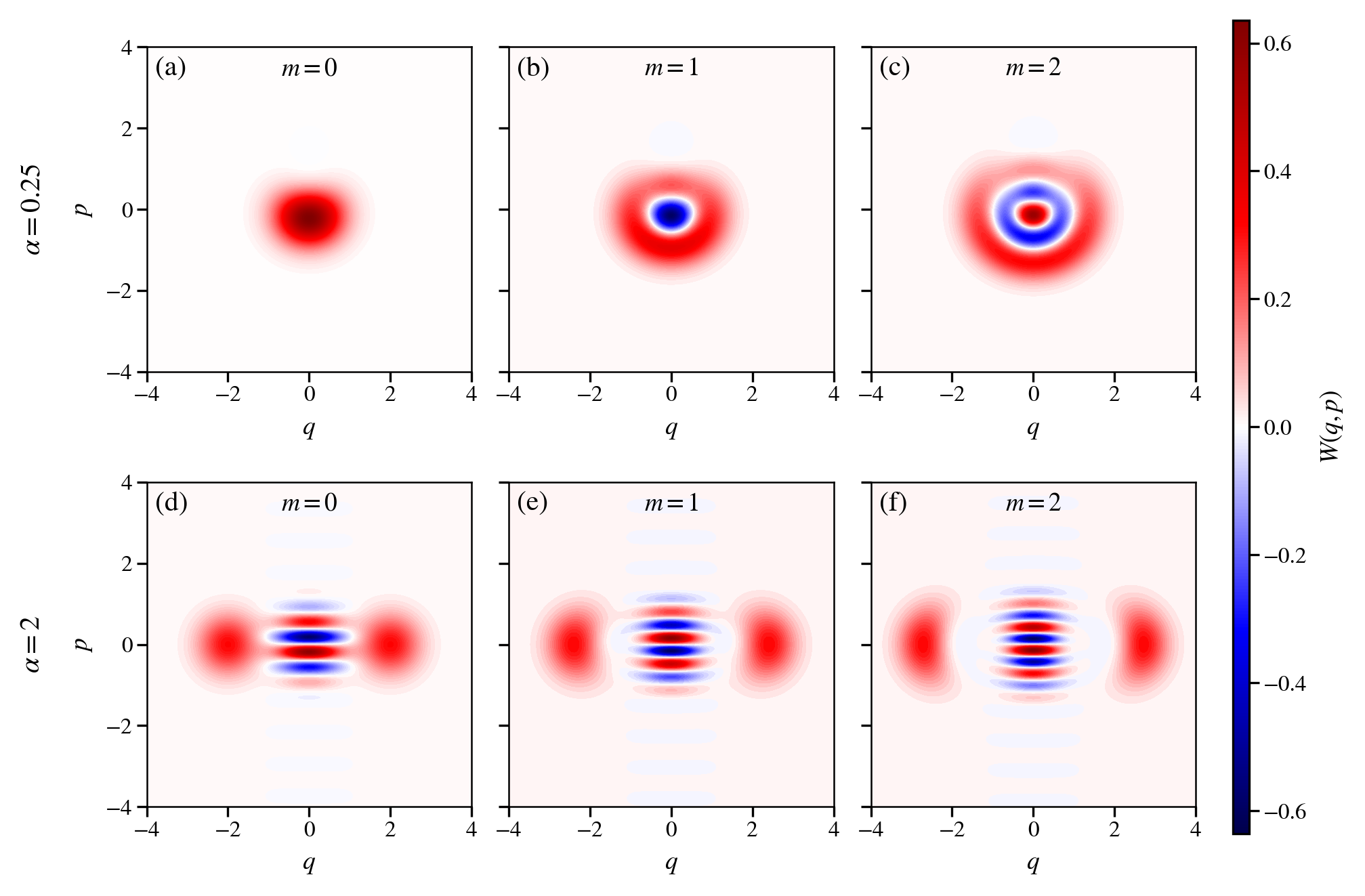}
  \caption{Wigner functions of the Yurke–Stoler state. Panels (a) and (d): original state. 
Panels (b) and (e): after a single-photon addition. Panels (c) and (f): a second photon addition returns the phase space 
distribution to the original configuration. The top row corresponds to $\alpha=0.25$ and the bottom row to $\alpha=2$.}
        \label{fig6}
\end{figure*}
As an example, in Figs. \ref{fig6}(a)-(f), using Eqs.~\eqref{3.1} and \eqref{3.2}, we present the Wigner function of the Yurke-Stoler state for different values of $m$ and both relatively weak ($\alpha=0.25$, first row) and strong ($\alpha=2$, second row) cat states. The Wigner functions clearly illustrate how photon addition reshapes the phase space structure of the original state: adding an odd number of photons introduces a $\pi$ phase shift with respect to the original superposition. A direct consequence of this effect is observed at the origin of the contour plot:  the Yurke-Stoler state is originally positive at $z=0$, but it becomes negative after a single-photon addition, signaling a transformation of the state parity. When a second photon is added (or, more generally, an even number of photons), the phase shift is neutralized, and the Wigner function recovers the original parity configuration, returning to a positive value at the center. This was already pointed out in Ref. \cite{Arman2021}. Moreover, when the amplitude $\alpha$ is increased, as shown in Figs. \ref{fig6}(d)–(f), the coherent components become more spatially separated in phase space. As a result, the oscillatory structure of the Wigner function becomes more pronounced, and each (unnormalized) photon-added coherent component tends to a Gaussian-like profile. In this regime, the negative (or positive) region around the origin becomes well isolated from the side lobes, making the effect of photon addition on the parity and interference fringes even more evident.

Another point worth mentioning is that the contour plot of the Wigner function in Fig. \ref{fig6} shows a clear deformation of phase space that, at first glance, resembles the compression usually associated with first-order squeezing. Our analysis in the previous section, however, shows that photon-added cat states do not exhibit first-order quadrature squeezing. Instead, they display amplitude-squared squeezing, a higher-order effect that does not appear in the standard quadrature variances yet still leaves a distinctive imprint on the phase space distribution. This illustrates how higher-order nonclassicality can reveal subtle, smooth nonclassical features that remain entirely hidden at lower orders.

\section{Experimental proposal for producing multi-photon-added cat states}\label{4.4}
Now, we propose two experimental schemes based on existing hardware enabling the heralded implementation of the $m$-photon-added cat states.

The first method relies on a $\lambda$-weak atom-light interaction between a two-level system prepared in an excited state and a photonic cat state generated externally~\cite{Ourjoumtsev2006, Simon2024,NeergaardNielsen2006}. Assuming a weak cat state to eliminate coherent population transfer between the atomic levels, the presence of the photonic field stimulates the atom emission into the cat state mode. Thus, by measuring the atom in its ground state at time $t$ shortly after the interaction, one heralds with probability $\lambda t\ll1$ the photon addition to the cat state. This can be beneficially realized in trapped-ion systems~\cite{Cetina2013, HAFFNER2008}, where the measurement of the deexcitation of the level system can be done efficiently by observing scattered light from this system \cite{Schmidt2005}. Additionally, this can be extended to $m$-photon addition by coupling several initialized ions with an identical cavity mode initialized with the cat state. Assuming perfect fidelity readout~\cite{Myerson2008} and measuring $m$-ions in the ground state shortly after the interaction, the $m$-times excited cat state heralding probability scales $(\lambda t)^{2m}$.

A similar idea can be extended to non-linear optics, where the matter qubits are replaced with $\chi^{(2)}$ non-linear crystals. Upon pumping with a strong pump laser pulses, these crystals spontaneously down-convert (SPDC) an entangled photon pair emitted into two modes (signal and idler) described by the operator $U\approx 1+ r a^\dagger_ia^\dagger_s$, with squeezing parameter $r\ll1$. Upon measuring one of these photons in the idler mode, we herald the presence of a single excitation in the signal mode, effectively emulating the stimulated emission~\cite{Barbieri2010}. If the non-linear crystal is seeded with an additional photonic state $|\psi\rangle$, the measurement of the idler photon heralds the generation of the single-photon added state $a^\dagger |\psi\rangle$. This mechanism was used earlier to generate single-photon added coherent~\cite{Zavatta2004,Zavatta2005} and thermal~\cite{Zavatta2007} states, or to generate directly cat states~\cite{Chen2024,Simon2024}. Using the cat state further as the seed state of $m$-cascaded SPDC crystals and heralding $m$-photons, we can again prepare $m$-times excited cat state with heralding probability $r^{2m}$ (assuming ideal detection efficiency).

\section{Conclusions}\label{5.5}
In this Letter, we have constructed multi-photon-added cat states and investigated several of their nonclassical signatures, including oscillations in the photon-number distribution, sub-Poissonian statistics, second-order squeezing, and Wigner function negativity. We have also proposed two feasible experimental schemes for generating such states with current technology, one based on atom–light interaction and the other on nonlinear optics. We found that photon addition induces a crucial trade-off: while even and Yurke–Stoler cat states become  sub-Poissonian, a desirable trait for quantum imaging, the same process completely suppresses first-order quadrature squeezing, whereas the odd cat state remains both sub-Poissonian and non-first-order squeezed. At the same time, photon addition introduces Hillery-type amplitude-squared squeezing, which is absent in the original cat states. Furthermore, the emergence of second-order squeezing, sub-Poissonian character of the field, and the lack of first-order squeezing are global properties that do not depend on the relative phase of the superposition. Consequently, if the initial cat state exhibits ordinary quadrature squeezing, this property is entirely lost after photon addition. The Wigner function also reveals that the parity of the superposition is inverted whenever an odd number of photons is added. Despite the absence of first-order quadrature squeezing, photon-added cat states display a clear deformation of phase space that resembles compression, an effect that originates from amplitude-squared squeezing, a smoother and more subtle form of nonclassicality that nevertheless leaves a mark on phase space.

\section*{Acknowledgments}
J. S. thanks Prof. Danieverton Moretti for insightful discussions.

\bibliographystyle{apsrev4-2}
\bibliography{ref}

\end{document}